\documentclass[journal]{IEEEtran}
\usepackage{cite}
\usepackage{graphicx} 
\usepackage{amsmath,amsfonts,amssymb}
\usepackage{dsfont}
\usepackage{url}

\usepackage{tikz}
\usepackage[caption=false,font=normalsize]{subfig}
\newcommand{\statetarget}{\mathbf{x}}

\newcommand{\stateRx}{\mathbf{r}}
\newcommand{\measure}{y}
\newcommand{\pos}{\mathbf{p}}
\newcommand{\posTx}{\pos}
\newcommand{\vel}{\mathbf{v}}
\newcommand{\timestep}{t}
\newcommand{\losgate}{g}
\newcommand{\adv}{\mathbf{a}}

\newcommand{\ctr}{\mathbf{u}}
\newcommand{\boundctr}{u_{\max}}
\newcommand{\seqctr}{\mathbf{U}}
\newcommand{\setctr}{\mathcal{U}}
\newcommand{\identitymat}{\mathbf{I}}
\newcommand{\zeromat}{\mathbf{0}}
\newcommand{\cov}{\mathbf{V}}
\newcommand{\var}{v}
\newcommand{\errcov}{\mathbf{C}}
\newcommand{\fim}{\mathbf{J}}
\newcommand{\statetransmat}{\mathbf{F}}
\newcommand{\ctrtransmat}{\mathbf{G}}

\DeclareMathOperator*{\argmin}{arg\,min}
\newcommand{\costfunc}{\mathcal{J}}

\newcommand{\weight}{\mathbf{W}}
\newcommand{\projpos}{\mathbf{q}}
\newcommand{\proj}{\Pi}
\newcommand{\norm}{\mathbf{n}}
\newcommand{\tdoatext}{\textsc{tdoa}}
\newcommand{\lostext}{\textsc{los}}

\newcommand{\linesegment}{\ell}
\newcommand{\noisevar}{\sigma^2}
\newcommand{\tdoavec}{\boldsymbol{\tau}}

\title{Obstacle-Aware Online Receiver Planning in Multistatic Ranging}
\author{Jingwei Hu, Dave Zachariah, Petre Stoica and Torbj\"orn Wigren
\thanks{All authors are with Dept. of Information Technology, Uppsala University, Sweden}}

\begin{document}
\maketitle

\begin{abstract}
Multistatic ranging with mobile receivers enables good tracking performance due to the combined adaptive sensor geometry. However, environments that contain signal obstructing obstacles require receiver trajectory planning to maintain line-of-sight (LOS) conditions with transmitters and the target of interest. In this letter, we develop a non-myopic receding-horizon framework for multistatic tracking. It uses convex collision-avoidance constraints and a control objective that focuses on maintaining good LOS signal conditions, taking into account future obstructions. We demonstrate the efficiency and tracking accuracy  of the method via a numerical experiment.
\end{abstract}


\section{Introduction}

Multistatic ranging systems enable geometric and path diversity of time-difference-of-arrival (\tdoatext{}) measurements by the  separation of transmitters and receivers \cite{li2008mimo,chernyak2018fundamentals}. Since the accuracy of estimated target states depends strongly on the sensor geometry, the receiver configuration design is central to localization and tracking. Mobile receivers, for example implemented by
UAV/UGV platforms \cite{chai2024cooperative}, can moreover be repositioned to maintain line-of-sight (\lostext{}) with transmitters and the target while actively improving the \tdoatext{} geometry as the target moves. Existing work improves estimation performance by optimizing the sensor geometry, either by static receiver placement \cite{fatima2024optimal} or adaptive receiver motion for multistatic target tracking \cite{zhan2010adaptive}. However, in cluttered environments, many methods either assume obstacle-free settings or handle occlusion via separate feasibility constraints. This assumption decouples visibility from informativeness and fails to anticipate occlusions caused by target motion and scenario geometry.

In cluttered environments, many  target-tracking approaches in robotics rely on policies that adapt sensor geometry over local or short-horizon motion, e.g., desired distances, bearings, and fields of view, while enforcing safety. For instance, in \cite{yao2015real} a sensing robot is driven towards a target while avoiding obstacles by combining Lyapunov guidance fields with an interfered fluid dynamical system. A separate line of work makes visibility explicit by optimizing geometric visibility surrogates to maintain \lostext{}: a differentiable visibility cost is introduced in \cite{wang2021visibility} that accounts for observation distance, angle and occlusion, and thereby optimizing trajectories for aerial tracking. Visibility is modelled probabilistically in \cite{gao2024probabilistic} under robot/target uncertainty using a belief-space metric to achieve non-myopic tracking in cluttered environments. While these approaches effectively maintain proximity to and visibility of the target, they are typically not explicitly design to maintain high estimation accuracy of the estimated target state.

These limitations motivates unified approach for multistatic active tracking in environments with obstructions. Our main contributions are:
\begin{itemize}
    \item robust obstacle-aware replanning: A non-myopic receding-horizon framework for multistatic tracking using convex collision-avoidance constraints.
    \item active non-\lostext{} mitigation: An objective for controlling receiver trajectories that focuses on maintaining good \lostext{} signal conditions when taking into account future obstructions.
\end{itemize}
The resulting method enables efficient, feasible motion control and target tracking.

\section{Problem Formulation}

We consider the problem of tracking a moving target in an environment with obstacles using $M$ mobile receivers and $N$ static transmitters. The goal is to accurately track the target by continuously planning the trajectory of receivers in $\mathbb{R}^d$, where $d=2$ or $3$.

\begin{figure}[!htb] 

\vspace{-1.5em}
    \centering
    \includegraphics[width=0.8\linewidth]{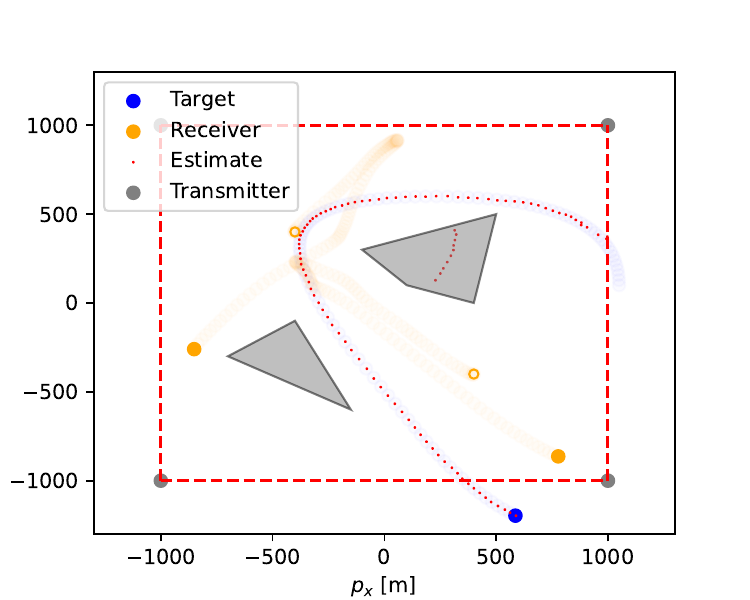}
    \caption{Target tracking in two-dimensions using $M=2$ mobile receivers in an environment with $O=2$ obstacles contained in convex polyhedra (gray shaded). The setup includes $N=4$ static transmitters. The receiver are initialized at $\circ$ and constrained to remain within the rectangular box. The sliding plan horizon is $k=15$ samples and the receiver accerlerations are limited to $u_{\max} =5 \text{m}/\text{s}^2$. (See Section~\ref{sec:experiments} for more details.)}
    \label{fig:two-obstacle-2D-case}
\end{figure}

The environment map contains static obstacles, indexed by $o \in
\{1, \dots, O\}$, which may block \lostext{} paths between transmitters, receivers, and the target. Here we follow a common approach in robotics by bounding the volume of each separable obstacle $o$ with a convex polyhedron, denoted $\Omega_o\subset \mathbb{R}^d$ \cite{lavalle2006planning,orthey2023sampling,paden2016survey}. An example configuration is illustrated in Figure~\ref{fig:two-obstacle-2D-case}. The map also contains the locations of the $N$ transmitters, denoted $\{ \posTx^n \}$. We assume that all nonobstructed points on the map can be reached.

The position and velocity of the \emph{target} are described in state-space form:
\begin{equation}
\statetarget_\timestep = \begin{bmatrix}
    \pos_\timestep\\
    \vel_\timestep
\end{bmatrix}, \quad \text{where } 
\statetarget_{\timestep}= \underbrace{\begin{bmatrix}
\identitymat & d\timestep \identitymat \\\zeromat & \identitymat\end{bmatrix}}_{\statetransmat} \statetarget_{\timestep-1}
+ \underbrace{\begin{bmatrix} \frac{1}{2} d\timestep^2\identitymat  \\ d\timestep \identitymat \end{bmatrix}}_{\ctrtransmat}  \adv_{\timestep}, 
\label{eq:targetdynamics}
\end{equation}
and $\adv_\timestep$ is an unknown acceleration. For the estimator used below, we will model the acceleration as a temporally white process  $\adv_\timestep \sim \mathcal{N}(\zeromat, \mathbf{Q})$ with a given covariance matrix $\mathbf{Q}$ that represents the physical limits of the target acceleration \cite{hu2025target}.

Next, consider receiver $m$, located at position $\pos^m$. It measures the \textsc{tdoa} between the signal arriving directly from transmitter $n$ and the reflected signal from the target. This \textsc{tdoa} measurement is modelled as
\begin{equation}
    \measure^{n,m} \sim \mathcal{N}\big(\tau_n(\pos,\pos^m),v_n(\pos,\pos^m)\big).
    \label{eq:measurment-scalar}
\end{equation}
where 
\begin{equation*}
\tau_n(\pos,\pos^m) =  \frac{1}{c} (\| \posTx^n - \pos \|_2 + \| \pos - \pos^m  \|_2) - \frac{1}{c} \| \posTx^n - \pos^m  \|_2 
\end{equation*}
and $c$ denotes the speed of light. The first term is the time of flight through the reflected path and the second term corresponds to the direct path. 

Under \lostext{}, the variance $v_n(\pos,\pos^m)$ depends on noises from  observing reflected and direct signals. Their variances are denoted $\noisevar_r$ and $\noisevar_d$, respectively, and are assumed to be calibrated. They can be distance dependent, and in the simplest case they can be taken as equal constants, i.e., $\noisevar_r = \noisevar_d$. In non-\lostext{} conditions, the measurement is taken to be \emph{uninformative}. We express this as
\begin{equation}
v_n(\pos,\pos^m) = \begin{cases}
    \noisevar_r + \noisevar_d, & \linesegment(\pos^n,\pos^m,\pos)\cap \Omega_o = \emptyset,\: \forall o  \\
    \infty, & \text{otherwise}
\end{cases}
\label{eq:variancemodel}
\end{equation}
where $\linesegment(\pos^n,\pos^m,\pos)$ is the union of the three line segments connecting the positions of the transmitter $n$, receiver $m$ and the target.  We note that identifiability of $\statetarget$ necessitates three signal paths -- transmitter-to-target, target-to-receiver, and transmitter-to-receiver -- to be in \lostext{} simultaneously. For a meaningful tracking scenario, we will therefore assume that all points on the map are in \lostext{} of \emph{at least one} static transmitter. In $d=2$, at least three measurements between transmitter $n$ and receiver $m$ are required since each measurement constrains the target location to an ellipse.

The position and velocity of receiver $m$ is described in state-space form:
\begin{equation}
\stateRx^m = 
\begin{bmatrix}
    \pos^m\\
    \vel^m
\end{bmatrix}, \; \text{where }
\stateRx^m_{\timestep}= \underbrace{\begin{bmatrix}
\identitymat & d\timestep \identitymat \\\zeromat & \identitymat\end{bmatrix}}_{\widetilde{\statetransmat}} \stateRx^m_{\timestep-1}
+ \underbrace{\begin{bmatrix} \frac{1}{2} d\timestep^2\identitymat  \\ d\timestep \identitymat \end{bmatrix}}_{\widetilde{\ctrtransmat}}  \ctr^m_{\timestep}.
\label{eq:receiverdynamics}
\end{equation}
The acceleration $\ctr^m_{\timestep}$ and states $\stateRx^m_{\timestep} $ are subject to \emph{physical} constraints:
\begin{equation}
\| \ctr_\timestep \|_2 \leq \boundctr \; \text{and} \; \{ \stateRx^m_\timestep : \stateRx_{\min} \leq \stateRx^m_\timestep \leq \stateRx_{\max}, \: \forall \timestep \}
\label{eq:boxconstraint}
\end{equation}
as well as \emph{collision-avoidance} constraints:
\begin{equation}
\big \{ \stateRx^m_\timestep :  [\identitymat_d \; \zeromat] \stateRx^m_{\timestep} \not \in  \Omega_o, \: \forall \timestep, o \big\}.
 \label{eq:obstacleconstraint}
\end{equation}

The goal of this paper is to \emph{jointly} determine the receiver control inputs $\{ \ctr^m_{\timestep+1}\} $ in \eqref{eq:receiverdynamics} so that the resulting \textsc{tdoa}  measurements at time $\timestep+k$ enable accurate estimation of the target state $\statetarget_{\timestep+k}$. This centralized strategy requires taking into account future signal obstructions by the obstacles in a tractlable manner.

\section{Method}

For target state tracking, we adopt the maximum likelihood framework in \cite{hu2025target}. The \textsc{tdoa} measurements \eqref{eq:measurment-scalar} from $M$ receivers at $\timestep$ are stacked into:
\begin{equation}
    \mathbf{\measure}_\timestep \sim \mathcal{N}\big(\tdoavec(\statetarget_{\timestep},\stateRx_\timestep),\cov(\statetarget_{\timestep},\stateRx_\timestep)\big),
\label{eq:measurement-vector}
\end{equation}
where $\stateRx_\timestep$ is a $2dM \times 1$  vector of receiver states and the diagonal matrix $\cov(\statetarget,\stateRx_\timestep)$ has diagonal elements given by \eqref{eq:variancemodel}. In addition to the measurements, we assume access to a linear predictor $\Check{\statetarget}_\timestep$ based on past data with a given error covariance matrix $\Check{\errcov}_\timestep$ using the model \eqref{eq:targetdynamics}. Specifically, we model the predictions as \cite{park2013gaussian}
\begin{equation}
\Check{\statetarget}_\timestep \sim \mathcal{N}( \statetarget_{\timestep}, \Check{\errcov}_\timestep )
\label{eq:prediction-vector}.
\end{equation}
By combining the measurements in \eqref{eq:measurement-vector} and the prior prediction in \eqref{eq:prediction-vector}, we  derive the maximum likelihood estimator of $\statetarget_\timestep$, which can be expressed as \cite{van1968detection,kay1993,hu2025target}:
\begin{equation}
\begin{split}
\widehat{\statetarget}_\timestep &= \argmin_{\statetarget}  \; \|\mathbf{y}_\timestep-\tdoavec(\statetarget,\stateRx_\timestep)\|^2_{\cov^{-1}(\statetarget,\stateRx_\timestep)}\\
    &+ \ln |\cov(\statetarget,\stateRx_\timestep)| + \|\statetarget - \Check{\statetarget}_\timestep\|^2_{\Check{\errcov}_\timestep}.
\end{split}
\label{eq:estimator}
\end{equation}
The predictor at any future time $\timestep+k$ is then   $\Check{\statetarget}_{\timestep+k} = \statetransmat^k \widehat{\statetarget}_\timestep$ \cite{kailath2000linear}.\footnote{Its error covariance matrix is updated as 
$\Check{\errcov}_{\timestep+k} :=  \statetransmat \errcov_{\timestep+k-1}\statetransmat^\top +   \ctrtransmat \mathbf{Q} \ctrtransmat^\top$, where $\errcov_{\timestep+k-1}=\Check{\errcov}_{\timestep+k-1}$ for $k>1$ or $\errcov_{\timestep}=(\fim_\timestep + \Check{\errcov}^{-1}_{t})^{-1}$ for $k=1$ using as an approximation of the ML-estimator's error covariance. See \eqref{eq:crb} and  \cite{kay1993}.} 

To determine the receiver inputs  $\ctr^m_{\timestep+1}$, we first stack them into a vector $\ctr_{\timestep+1}$. At time $\timestep$, the planned trajectories of all $M$ receivers up to time $\timestep +k$ are determined by
$$\seqctr_{\timestep} := \{\ctr_{\timestep+1},\dots,\ctr_{\timestep+k}\}.$$
We use a sliding-window approach to plan the trajectories:
\begin{equation}
\begin{split}
\seqctr^*_{\timestep} &= \argmin_{\seqctr_\timestep \in \setctr \cap \setctr_c}  \costfunc(\seqctr_{\timestep}),
\end{split}
\label{eq:controlproblem}
\end{equation}
where $ \setctr$ and $\setctr_c$ are sets corresponding to constraints \eqref{eq:boxconstraint} and \eqref{eq:obstacleconstraint}, respectively. The criterion $\costfunc(\seqctr_{\timestep})$ is designed to incentivize input sequences that result in good tracking of the target at time $\timestep+k$ starting from current estimates. At each time step, the plan \eqref{eq:controlproblem} is recomputed and $\ctr^*_{t+1}$ is implemented as a control input.

Next, we derive tractable formulations of $\setctr$ and $\setctr_c$. We then turn to formulating a criterion $\costfunc(\seqctr_{\timestep})$ that incentivizes receiver trajectories that maintain good tracking while taking into account possible obstructions by obstacles $o \in
\{1, \dots, O\}$.

\subsection{Formulation of tractable constraints in \eqref{eq:controlproblem}}\label{sec:obstacle-constraints}

We begin by formulating the collision-avoidance constraints \eqref{eq:obstacleconstraint}. That is, a feasible space for each receiver $m$ that excludes all obstacles. Let $\pos$ be any position in $\mathbb{R}^d$. Then the closest point on an obstacle $o$ to $\pos$ is given by
\begin{equation}
 \proj_o(\pos) := \argmin_{\projpos \in\Omega_o}\| \pos- \projpos \| 
\label{eq:projection}
\end{equation}
and let
\begin{equation}
\norm_o(\pos):=\frac{\proj_o(\pos)-\pos}{\|\proj_o(\pos)-\pos\|}
\label{eq:normal}
\end{equation}
be the unit vector pointing from $\pos$ to the projected point. Using this notation, we can express the half-space that \emph{contains} receiver $m$ located at $\pos^m$ and \emph{excludes} obstacle $o$ as:
\begin{equation}
\big\{\pos \in \mathbb{R}^d \: : \: (\projpos^{o,m}-\pos)^\top \norm^{o,m} >0 \big\},\label{eq:supporting-hp}
\end{equation}
where $\projpos^{o,m} = \proj_o(\pos^m)$ and $\norm^{o,m} =\norm_o(\pos^m)$ jointly determine an orthogonal supporting plane of $\Omega_o$.  The intersection of excluding halfspaces \eqref{eq:supporting-hp} across \emph{all} obstacles $o=\{ 1,\dots, O\}$ thus expresses the constraint \eqref{eq:obstacleconstraint}. 

The computation of $\projpos^{o,m}$ is a quadratic programming problem  \cite{boyd2004convex}, which for $d=2$ or $3$ can be solved directly in closed form \cite{ericson2004real}. For tractability, we relax the intersection of halfspaces \eqref{eq:supporting-hp} across the planning horizon by considering the space that is ensured to be feasible at time $\timestep$. That is,
\begin{equation}
\setctr_c = \Big\{ \seqctr_{\timestep} : \mathbf{A}^m_t [\identitymat_d \: \zeromat] \stateRx^m_{\timestep+i} \leq \mathbf{b}^m_t, \: \forall  m,i=1, \dots,k \Big\},
\label{eq:obstacleconstraint_alt}
\end{equation}
where 
\begin{equation}
    \mathbf{A}^m_t = \begin{bmatrix}
        \norm^{1,m}_t\\
        \vdots\\
        \norm^{O,m}_t\\
    \end{bmatrix} \; \text{and} \; \mathbf{b}^m_t = \begin{bmatrix}
        (\norm^{1,m}_t)^\top \projpos^{1,m}_t\\
        \vdots\\
        (\norm^{O,m}_t)^\top \projpos^{O,m}_t\\
    \end{bmatrix}
\end{equation}
Using \eqref{eq:receiverdynamics}, $\stateRx^m_{\timestep+k}$ in \eqref{eq:obstacleconstraint_alt} can be expressed as
\begin{equation}
   \stateRx^m_{\timestep+k} =  \widetilde{\statetransmat}^{k}\stateRx_{\timestep}^m + \sum^{k}_{i=1}\widetilde{\statetransmat}^{i-1}  \widetilde{\ctrtransmat} \ctr^m_{\timestep+i}.
\label{eq:receivertrajectory}
\end{equation} 
Thus $\setctr_c$ is a relaxed local convex collision-free set at $\timestep$, formed by \emph{linear} inequality constraints with respect to $\seqctr_\timestep$. The  $\setctr_c $ is recomputed at each $\timestep$, and can thus change as a receiver moves across a polyhedron. With large enough $k$, the relaxation remains feasible, producing a sequence of overlapping collision-free sets which enable collision-free paths across the map.
A related idea appears in the convex feasible set algorithm of \cite{liu2018convex},
where nonconvex collision-avoidance constraints are handled by constructing
convex inner approximations through a sequence of supporting hyperplanes.



We can also express \eqref{eq:boxconstraint} as 
\begin{equation*}
   \setctr = \Big\{ \ctr_{\timestep+i}^m : \| \ctr^m_{\timestep+i} \|_2 \leq \boundctr, \stateRx_{\min} \leq \stateRx^m_{\timestep+i} \leq \stateRx_{\max} \: \forall m,i \Big \}. 
\end{equation*}
The norm-bound can be relaxed using the $\ell_\infty$-norm and thus we relax \eqref{eq:boxconstraint} as:
\begin{equation*}
 \widetilde{\setctr} = \Big\{ \ctr_{\timestep+i}^m : \| \ctr^m_{\timestep+i} \|_\infty \leq \frac{\boundctr}{\sqrt{d}}, \stateRx_{\min} \leq \stateRx^m_{\timestep+i} \leq \stateRx_{\max} \: \forall m, i \Big \} .
\end{equation*} 
Then both constraint sets $\widetilde{\setctr}$ and $\setctr_c$ can be described by linear inequalities, which allows for efficient implementation when solving \eqref{eq:controlproblem}.

\subsection{Formulation of criterion in \eqref{eq:controlproblem}}


At time $\timestep$, we plan the receiver trajectories \eqref{eq:controlproblem} that yield an informative future measurement $\measure_{\timestep+k}$. Using this measurement, the resulting tracking error is measured in common units:
\begin{equation}
\mathbb{E}\Big[ \| \statetarget_{\timestep+k} - \widehat{\statetarget}_{\timestep+k} \|^2_{\mathbf{W}}\Big], \: \text{where } \mathbf{W} = \begin{bmatrix}
\identitymat & \zeromat \\
\zeromat & dt \identitymat
\end{bmatrix}.
\label{eq:mse}
\end{equation}
Our objective is to continually steer $\stateRx_{\timestep}$ to reduce the future tracking error \eqref{eq:mse} which requires maintaining \lostext{} conditions. The error is lower bounded via the expected Cramér–Rao inequality (CRB) \cite{van1968detection, kay1993, stoica2005spectral}:
\begin{equation}
\mathbb{E}\Big[ \| \statetarget_{\timestep+k} - \widehat{\statetarget}_{\timestep+k} \|^2_{\mathbf{W}}\Big] \geq \mathbb{E}[f( \statetarget_{t+k},\stateRx_{t+k} )],
\label{eq:expected-crb}
\end{equation}
where
\begin{equation}
f( \statetarget_{t+k},\stateRx_{t+k} ) = \text{tr}\left\{ \weight \big( \fim(\statetarget_{t+k}, \stateRx_{t+k}) + \Check{\errcov}^{-1}_{t+k} \big)^{-1} \right\}.
\label{eq:crb}
\end{equation} 
Here $\fim$ denotes the Fisher Information Matrix\cite{kay1993,stoica2005spectral} 
\begin{equation}
\begin{split}
[\fim(\statetarget,\stateRx)]_{ij} &= \frac{\partial{\tdoavec(\statetarget,\stateRx)}}{\partial{x}_i}^\top \cov^{-1}(\statetarget,\stateRx) \frac{\partial{\tdoavec(\statetarget,\stateRx)}}{\partial{x}_j}\\ &+ \frac{1}{2}\text{tr}\left\{\cov^{-1}(\statetarget,\stateRx)\frac{\partial{\cov(\statetarget,\stateRx)}}{\partial{x}_i}\cov^{-1}(\statetarget,\stateRx)\frac{\partial\cov(\statetarget,\stateRx)}{\partial{x}_j}\right\},
\end{split}
\label{eq:fim}
\end{equation} 
where \lostext{} conditions enter into the covariance matrix $\cov$ via \eqref{eq:variancemodel}. Using \eqref{eq:receivertrajectory}, the bound on the anticipated tracking error \eqref{eq:expected-crb} can be expressed as
\begin{equation}
\mathbb{E}[f( \statetarget_{t+k},\stateRx_{t+k} )] \equiv \mathbb{E}[ f(\statetarget_{t}, \adv_{t+1}, \dots, \adv_{t+k}, \seqctr_\timestep )].
\label{eq:anticipated_crb}
\end{equation}
This is a suitable control objective since it incentivizes receiver trajectories that can maintain good signal conditions with the target and the transmitters. However, the noise model \eqref{eq:variancemodel} renders the Fisher information \eqref{eq:fim} discontinuous in the receiver trajectories. 

To circumvent this computational challenge, we modify the  information obtained by a future measurement from transmitter $n$ to vary with the \lostext{} \emph{margins} as defined next. Consider a space in which the target and transmitter $n$ are \emph{jointly} ensured to be in \lostext{}.  Specifically, in the intersecting space 
$$\mathcal{L}(\pos) = \bigcap_{o=1}^O\big \{\mathbf{s}\in\mathbb{R}^d\: : \: (\proj_o(\pos)-\mathbf{s})^\top \norm_o(\pos) \ge 0\big\},$$
\lostext{} is guaranteed with respect to location $\pos$. If the intersection between $\mathcal{L}(\pos)$ and $\mathcal{L}(\pos^n)$ is nonempty at $\timestep+k$, this ensures a space in which all points are in \lostext{} with the target and transmitter $n$. At the same time it is necessary for the target and transmitter to be in \lostext{} with each other. These conditions are summarized  using the indicator function
\begin{equation*}
L(\pos, \pos^n) = \mathds{1}\big\{ \mathcal{L}(\pos)\cap\mathcal{L}(\pos^n) \not=\emptyset  \big\} \mathds{1}\big\{ \linesegment(\pos^n,\pos)\cap \Omega_o = \emptyset,\: \forall o \big\}
\end{equation*}
that is independent of any receiver trajectory. If $L(\pos, \pos^n) = 0$, \lostext{} cannot be ensured  and corresponding future measurement is taken to be uninformative. Thus, in lieu of \eqref{eq:variancemodel} we use:
\begin{equation}
\begin{split}
\widetilde{\var}_n(\pos,\pos^m) = \begin{cases} 
\frac{\noisevar_r}{{\losgate}_{\lambda_r}(\pos,\pos^m)} + \frac{\noisevar_d}{{\losgate}_{\lambda_d}(\pos^n,\pos^m)},& L(\pos, \pos^n) = 1 \\
\infty, & L(\pos, \pos^n)= 0,
\end{cases}
\end{split}
\label{eq:variancemodel_relax}
\end{equation}
where $\losgate_{\lambda}(\pos,\pos')$ is a continuous function that increases with the \lostext{} margins between $\pos$ and $\pos'$. Specifically, we use the sigmoid function
\begin{equation}
        \losgate_{\lambda}(\pos,\pos') = \left[ 1 + \exp\left(-\frac{\phi(\pos,\pos')}{\lambda}\right) \right]^{-1} \in (0,1)
\label{eq:soft-losindicator}
\end{equation} 
where $\lambda > 0$ is a parameter to control its sharpness and $\phi(\pos,\pos')$ is defined in the following way. Consider the plane that separates point $\pos'$ from obstacle $o$  as defined by \eqref{eq:projection} and \eqref{eq:normal}. Then 
$(\proj_{o}(\pos')-\pos)^\top {\norm_{o}(\pos')}$ is the signed \emph{margin} of position $\pos$ to this separating plane. That is, a positive margin ensures \lostext{} between $\pos$ and $\pos'$ with respect to obstacle $o$. In contrast, a negative margin provides no certificate of \lostext{}. Then the the \lostext{}-margin function \eqref{eq:soft-losindicator} is determined by the minimum margin
$$\phi(\pos,\pos'):=\min_{o\in \{1,\dots,O\}} (\proj_{o}(\pos')-\pos)^\top{\norm_{o}(\pos')}.$$ 
In other words, when \lostext{} conditions are ensured, the informativeness of a future measurement of receiver $m$ with transmitter $n$  increases with the \lostext{} margins. By setting $\lambda_r \gg \lambda_d$ in \eqref{eq:variancemodel_relax}, the receiver's \lostext{} margins to the target is taken to be more informative than an equivalent margin to the transmitter. As a result, preference is given to designs that prioritize \lostext{} margins with the target.

Let $\widetilde{f}( \statetarget_{t+k},\stateRx_{t+k} )$ denote \eqref{eq:crb} using model \eqref{eq:variancemodel_relax}. We approximate the expectation in \eqref{eq:anticipated_crb} using deterministic sigma-point sampling  of the target trajectory $(\statetarget_{\timestep}, \adv_{t+1}, \dots, \adv_{t+k})$, 
$\adv_t$ denotes the target acceleration in model \eqref{eq:targetdynamics}. A trajectory is generated by propagating the plug-in state estimate  $\widehat{\statetarget}_{\timestep}$ and its error covariance $\widehat{\errcov}_\timestep$ through the system dynamics \cite{candy2016bayesian,hu2025target}. We finally arrive at a  sampled-based criterion:
\begin{equation}
\begin{split}
\costfunc(\seqctr_{\timestep}) = \sum^{2(2+k)d+1}_{\ell=0} \widetilde{f}(\widehat{\statetarget}^{(\ell)}_{t}, \adv^{(\ell)}_{t+1}, \dots, \adv^{(\ell)}_{t+k}, \seqctr_\timestep ).
\end{split}
\label{eq:criterion_approx}
\end{equation} 
The computational demand scales linearly with $k$ but the problem is tractable due to the independence of the terms in the sum and the possibility of efficient (sub)gradient computations. The criterion \eqref{eq:criterion_approx} can now be used in \eqref{eq:controlproblem}.

\section{Numerical Experiment}
\label{sec:experiments}

The proposed method is evaluated using synthetic data. Each control update $\ctr^m_{t+1}$ is obtained by solving \eqref{eq:controlproblem} as a sliding-window with a standard interior-point method\cite{wachter2006implementation}. For illustration, we consider a tracking scenario with $M=2$ mobile receivers and $O=2$ obstacles contained in polygons as shown in Figure~\ref{fig:two-obstacle-2D-case}. $N=4$ static transmitted are located at the corners of a rectangle. Here $\sigma_r = \sigma_d = 3/c$ [s] and the sampling time is $dt=1$ [s]. For a target with a maximum acceleration of 5 $[\text{m/}\text{s}^2]$, we set $\mathbf{Q} = \frac{5^2}{3}\mathbf{I}_d$. The effects of varying the horizon length \(k\) and the maximum target acceleration on tracking performance in nonobstructing settings were already studied in \cite{hu2025target}.

We used a horizon of $k = 15$ samples, which depends on sampling period and the velocity limits of the receivers. In \eqref{eq:variancemodel_relax} we set $\lambda_d = 1$ and $\lambda_r = 10$. The experiments run on a standard laptop with nonoptimized code, requiring approximately two seconds to compute each control update \eqref{eq:controlproblem}. The tracking is initialized with no prior knowledge of the target and we consider two baselines: First, static receivers placed to have \lostext{} with all transmitters and second mobile receivers that implement \eqref{eq:controlproblem} but assume there are no \emph{signal} obstructions in the environment.

\begin{figure}[!htb]
    \centering
    \includegraphics[width=0.75\linewidth]{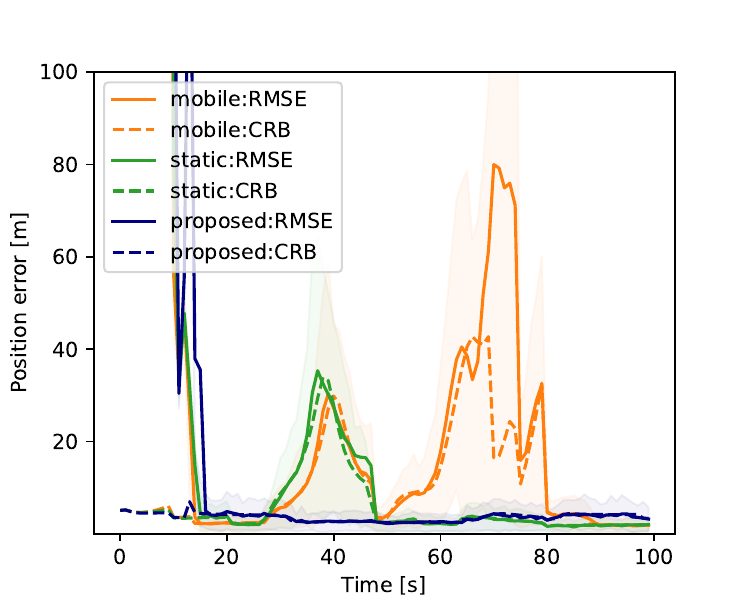}
    \caption{Tracking performance of the obstacle-aware receiver-planning method and two baseline methods in the map shown in Figure~\ref{fig:two-obstacle-2D-case}. Evaluation using 100 Monte Carlo simulations. The solid curve denotes RMSE, and the dashed line shows the Cramér-Rao bound (CRB). The shaded regions indicate the 5\%-95\% quantile range of estimation errors. }
    \label{fig:err_compare}
\end{figure}

The tracking errors are shown in Figure~\ref{fig:err_compare}. We see that  well-placed static receivers can track the target accurately as long as it remains in \lostext{}. When this signal condition fails, however, the CRB increases sharply and, correspondingly, the tracking errors increase significantly as well as exhibit large variability across runs. In contrast, mobile receivers that fail to take future \lostext{} obstructions into account can result in even more significant periods of large tracking errors. The proposed obstacle-ware method shows that once the estimator locks onto the the target, accurate tracking is maintained throughout its maneuver in between obstacles shown in Figure~\ref{fig:two-obstacle-2D-case}. In other words, the proposed method plans collision-free receiver trajectories that proactively maintain \lostext{} whenever feasible. An additional experiment with a different target trajectory is provided in the Supplementary Material.

\section{Conclusion}
We addressed the problem of target tracking using mobile receivers in environments with signal-obstructing obstacles. We proposed a non-myopic, receding-horizon framework that couples convex collision-avoidance constraints with an obstacle-aware control objective to maintain \lostext{} among transmitters, receivers, and the target. Our experiments illustrate that collision avoidance is insufficient for tracking in cluttered environments, as it fails to anticipate occlusions induced by target motion and scene geometry. By explicitly incorporating \lostext{} conditions into the planning horizon, the proposed method enables proactive receiver reorientation to maintain informative measurements and recover \lostext{} following interruptions. 

Future extensions include scaling the framework to multi-target scenarios, and incorporating richer \lostext{} metrics to model gradual sensing degradation under partial occlusion.

\bibliographystyle{IEEEtran}
\bibliography{ref}

\end{document}